%% file: main.tex
\documentclass[times,final]{elsarticle}

\usepackage{jcomp}
\usepackage{framed,multirow}

\usepackage{amssymb}
\usepackage{latexsym}
\usepackage{amsmath}

\usepackage{url}
\usepackage{xcolor}
\definecolor{newcolor}{rgb}{.8,.349,.1}

\journal{Journal of Computational Physics}

\begin{document}

\verso{Kai Leong Chong \textit{et al.}}

\begin{frontmatter}

\title{Multiple-resolution scheme in finite-volume code for active or passive scalar turbulence}%

\author[1]{Kai Leong \snm{Chong}\corref{cor1}\fnref{fn1}}
\fntext[fn1]{Present address: Physics of Fluids Group and Max Planck Center Twente, MESA+Institute, J. M. Burgers Centre for Fluid Dynamics, University of Twente, 7500 AE Enschede, Netherlands}  
\author[1]{Guangyu \snm{Ding}}
\author[1]{Ke-Qing \snm{Xia}\corref{cor2}}
\cortext[cor1]{Corresponding author: 
  Email.: stevenphycuhk@gmail.com}
  \cortext[cor2]{Corresponding author: 
  Email.: kxia@cuhk.edu.hk}

\address[1]{Department of Physics, The Chinese University of Hong Kong, Shatin, Hong Kong, China}

\received{x May 2018}
\finalform{x x 2018}
\accepted{x x 2018}
\availableonline{x x 2018}
\communicated{xx}

\begin{abstract}
In scalar turbulence it is sometimes the case that the scalar diffusivity is smaller than the viscous diffusivity. The thermally-driven turbulent convection in water is a typical example.  In such a case the smallest scale in the problem is the Batchelor scale $l_b$, rather than the Kolmogorov scale $l_k$, as $l_b = l_k/Sc^{1/2}$, where Sc is the Schmidt number (or Prandtl number in the case of temperature). In the numerical studies of such scalar turbulence, the conventional approach is to use a single grid for both the velocity and scalar fields. Such single-resolution scheme often over-resolves the velocity field because the resolution requirement for scalar is higher than that for the velocity field, since $l_b<l_k$ for $Sc>1$. In this paper we put forward an algorithm that implements the so-called multiple-resolution method with a finite-volume code. In this scheme, the velocity and pressure fields are solved in a coarse grid, while the scalar field is solved in a dense grid. The central idea is to implement the interpolation scheme on the framework of finite-volume to reconstruct the divergence-free velocity from the coarse to the dense grid. We demonstrate our method using a canonical model system of fluid turbulence, the Rayleigh-B\'enard convection. We show that, with the tailored mesh design, considerable speed-up for simulating scalar turbulence can be achieved, especially for large Schmidt (Prandtl) numbers. In the same time, sufficient accuracy of the scalar and velocity fields can be achived by this multiple-resolution scheme. Although our algorithm is demonstrated with a case of an active scalar, it can be readily applied to passive scalar turbulent flows.
\end{abstract}


\end{frontmatter}

\input{introduction}
\input{singleres}
\input{multires}
\input{validation}
\input{conclusion}

\end{document}

%% file: introduction.tex
\section{Introduction}

The advection of scalar fields by turbulent flows is a ubiquiotus natural process in daily life. An example of scalar turbulence is thermally-driven convection \citep{rayleigh1916}, thermohaline circulation \citep{apel1987principles} where salinity also plays a critical dynamical role, pollutants dispersion in atmosphere \citep{arya1999air}, and the spreading of chemicals, etc. In general, scalar fields can be classified into two kinds, i.e. the active and passive scalars,  depending on whether it can cause dynamical effect to the turbulent flow. Active scalars have back effect to the turbulent flow itself, for example, the buoyancy force generated by variations in the temperature field can influence the motion of fluid. In contrast, passive scalars are those having negligible effect on the fluid flow or they are introduced to the flow independently, for example the chemical dyes for flow visualization purpose.

Extensive works have been done through experiments in the study of scalar turbulence \citep{warhaft2000passive,shraiman2000scalar}. However, the role of numerical simulation is inevitably important in fluid mechanics research since complete flow and scalar fields can only be obtained from simulations. Through the advancement of high-performance computers, Direct Numerical Simulations (DNS) becomes a powerful tool to tackle the problem of scalar turbulence \citep{moin1998direct,pope2001turbulent}, which does not involve any small-scale modelling or parametrization. The idea of DNS is based on the turbulent energy cascade process. In order to accurately simulate the turbulent flow, one has to capture all the relevant spatiotemporal scales introduced by the cascade process. For three-dimensional turbulent flows, the energy transfer picture is that the kinetic energy is transferred successively from large scales to the small scales until the molecular viscosity becomes effective. This physical picture of the so-called direct energy cascade is firstly proposed by Richardson \citep{richardson2007weather} in  the 1920s. A theoretical framework is further proposed by Kolmogorov who suggested that, for flow at high enough Reynolds number, the kinetic energy spectrum should exhibit $k^{-5/3}$ over certain range of the spectrum known as the inertial subrange \citep{frisch1995turbulence,pope2001turbulent}, where $k$ is the wavenumber. However, at the scales below this subrange, the effect of viscous dissipation sets in and this length scale is known as the Kolmogorov length scale. Using dimensional analysis, one can obtain the expression of Kolmogorov length scale as $l_k=(\nu^3/\epsilon)^{1/4}$, where $\nu$ and $\epsilon$ are the kinematic viscosity of the fluid and the mean viscous dissipation rate respectively. Likewise, a corresponding length scale for scalar field can be obtained which is known as the Batchelor length scale given by $l_b=(\nu D^2/\epsilon)^{1/4}$ where $D$ is the scalar diffusivity. It can be seen that the Kolmogorov and the Batchelor length scales are related by $l_b=l_k/Sc^{1/2}$ where $Sc=\nu/D$ is the Schmidt number (in the case of temperature field Schmidt number can be replaced by Prandtl number).

In the DNS study of scalar turbulence, the properties of scalar substance can be done either by Eulerian or Lagrangian approaches. The latter one usually encounter the situation of localized sources where the certain amount of substance can be tracked individually by the  equations of  motion. The Eulerian approach provides a different perspective to the problem and has been widely used in the community for studying scalar turbulence. For this approach, the scalar field represents the concentration of certain scalar "substance", and its time evolution is governed by the advection-diffusion equation. The present paper focuses on the Eulerian approach for studying scalar turbulence. In order to accurately simulate the scalar field as well as the turbulent flow itself by DNS, an important criterion is to apply a mesh that resolves $l_b$ for the scalar field and $l_k$ for the velocity field. In the traditional approach, i.e. the single-resolution scheme, the minimum among $l_b$ and $l_k$ has to be considered in designing the mesh. For the case of large Sc, for example, thermal convection in water at $T=42^{\circ}C$ for which $l_b=0.5l_k$, the mesh for the velocity field is actually over-resolving. The unnecessary amount of grid points for velocity field put a burden on the performance of the code and computational resources, especially the velocity and pressure solver account for the majority of the computation cost.

The above discussion makes it clear the need for developing some more efficient schemes in simulating scalar turbulence, especially for large values of Sc. Previous works mainly focused on solutions for the situation of passive scalars. For instance, Verma \& Blanquart \citep{verma2013filtering} had proposed the velocity-resolved scalar-filtered approach which uses a kind of filtering in the viscous-convection subrange similar to the concept of Large Eddy Simulation. Gotoh \textit{et al.} \citep{gotoh2012spectral} had proposed the hybrid method for solving the passive scalar isotropic turbulence where the incompressible Navier-Stokes equation governing the velocity field is solved through the spectral method whereas the passive scalar field is solved in a denser grid by combined compact difference scheme. One of the few studies aimed at improving simulations with active scalars is by Ostillia-Monico \textit{et al.} \citep{ostillamonico2015jcp}.  They had proposed a multiple-resolution scheme where the coupled velocity and temperature fields are solved separately on the coarse and the dense meshes. The main idea is that the velocity field has to be reconstructed to match the dense meshes for every time step with divergence-free condition being guaranteed. In their finite-difference code, the velocity field cannot be simply interpolated but interpolation is made by their gradient. However, this approach with finite-difference algorithm can not be directly applied to finite-volume algorithm.

In this paper, we propose a novel way to reconstruct the divergence-free velocity from coarse to dense mesh with finite-volume algorithm, which is completely different from the method adopted to finite-difference code \citep{ostillamonico2015jcp}. We demonstrate our multiple-resolution scheme with finite-volume algorithm based on a code designed for simulating the Rayleigh-B\'enard (RB) convection. The code is called \textit{CUPS}. It was developed by Schmitt \& Friedrich \citep{Schmitt82} for simulating pipe and channel flows and was written in Fortran 77 standard. Later, the code was extended to study thermal convection by Shishkina \& Wagner \citep{shishkina2005fourth,shishkina2007fourth,shishkina2007boundary} in cylindrical geometry, and by Kaczorowski, Shishkin, Shishkina and Wagner \citep{kaczorowski2008nrnefm6} in Cartesian geometry. Next, Chong \& Kaczorowski adopted the pencil domain decomposition for MPI for the code \cite{kaczorowski2013jfm,kaczorowski2014jfm},  and also rewrote the code in Fortran 90 standard, which became the present \textit{CUPS} code. The Rayleigh-B\'enard convection, a fluid layer heated from below and cooled from above, is a typical example of turbulent flows with an active scalar  \citep{ahlers2009rmp,lohse2010arfm,chilla2012epj,xia2013taml}. We remark that with the scheme proposed here, the multiple-resolution algorithm can be readily applied to passive scalar turbulent flows.

The rest of the paper is organized as follows: Sec. 2 is a brief description of the details of the \textit{CUPS} code, including a description of the algorithm, the spatial and temporal discretization schemes and parallelization strategy; 
Sec. 3 contains description on the grid design in multiple-resolution scheme, the idea of reconstructing the divergence-free velocity field from coarse to dense mesh, the upscale integration of the temperature field and the routine of the multiple-resolution scheme. Sec. 4 presents a verification of the multiple-resolution code using the system of Rayleigh-B\'enard convection and performance profiling of the code; Sec. 5 is the Conclusion.

%% file: singleres.tex
\section{Description of \textit{CUPS} code}
\subsection{Governing equations}
The non-dimensional Navier-Stokes equations for incompressible thermal convection under Boussinesq approximation can be written as,
\begin{equation}
\label{equ:momentum_equation}
\partial_t u_i + u_j \partial_j u_i = -\partial_i P + T \delta_{iz} + \sqrt{\frac{Pr}{Ra}}\partial_j\partial_j u_i,
\end{equation}
\begin{equation}
\label{equ:heat_equation}
\partial_t T + u_i \partial_i T =  \sqrt{\frac{1}{RaPr}}\partial_i^2 T,
\end{equation}
\begin{equation}
\label{equ:continuity}
\partial_i u_i=0,
\end{equation}
where $\vec u$, $P$, $T$ are the dimensionless velocity, pressure and temperature, respectively. Non-dimensionality of the equaitons is made by using the cell height $H$, the free-fall velocity $(\alpha g \Delta T H)^{1/2}$, the free-fall time scale $(H/\alpha g \Delta T)^{1/2}$ and the top-and-bottom temperature difference $\Delta T$. Here $\alpha$ and $g$ are the thermal expansion coefficient and the gravitational acceleration. For RB convection, the dynamics of fluid flow is governed by the Rayleigh number $Ra=\alpha g \Delta T H^3/\nu \kappa$ and the Prandtl number $Pr=\nu/\kappa$ for a given cell geometry, where $\nu$ and $\kappa$ are respectively the kinematic viscosity and thermal diffusivity of the fluid.

\begin{figure}[h!]
	\centering
	\includegraphics[width=.6\textwidth]{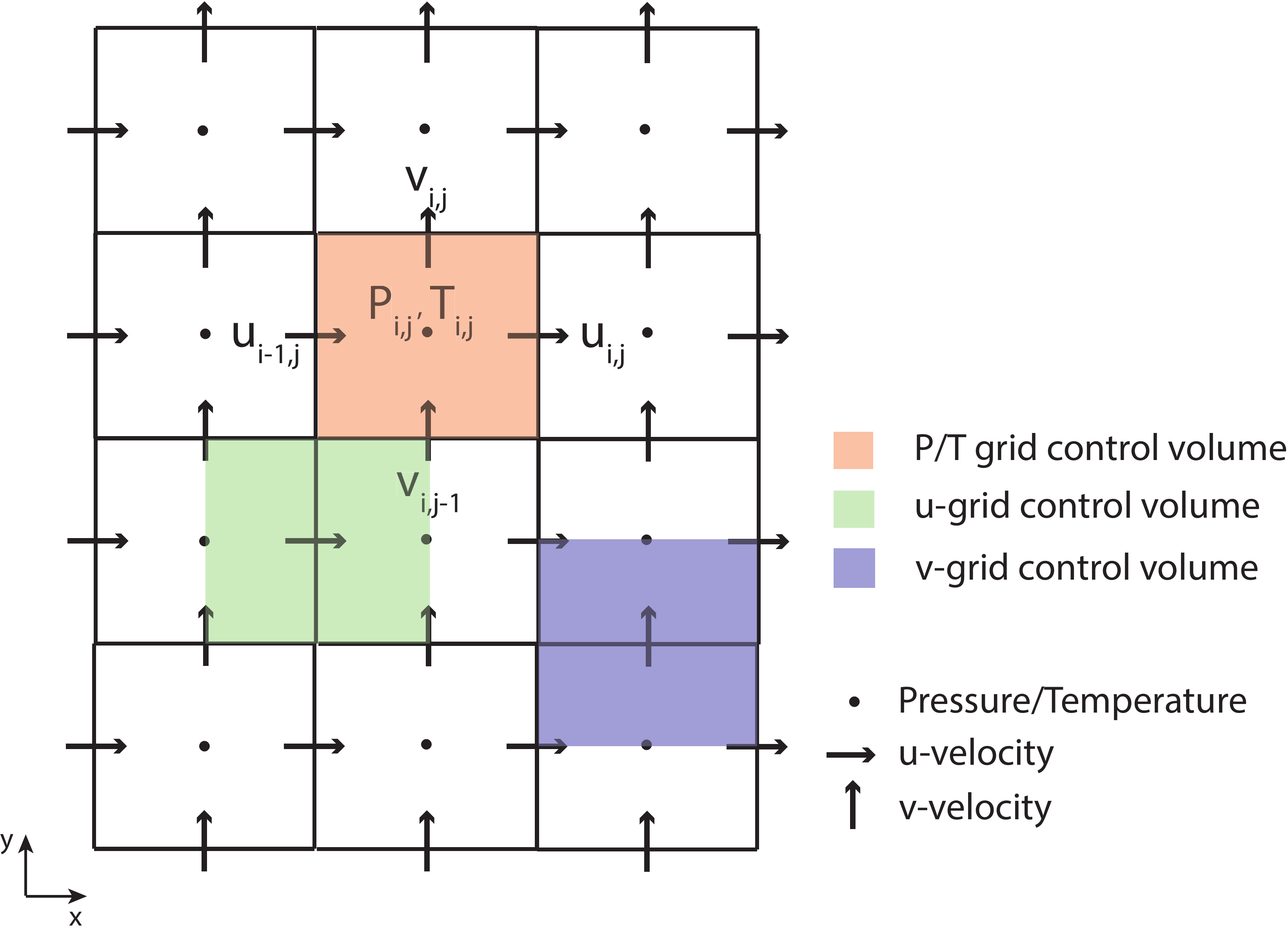}
	\caption{Schematic of the staggered grid}
	\label{fig:stagger}
\end{figure}

\subsection{Discretization}
Our code applies the finite-volume (FV) method to solve the momentum and heat equations. FV method divides the domain into a finite number of control volumes (CVs), such that the global conservation laws hold inherently since the flux across the faces of CVs will be balanced by each other. The physical quantities are defined over the CV instead of the computational node. The momentum and temperature equations is represented in integral form, by integrating Eqs. (\ref{equ:momentum_equation}) and (\ref{equ:heat_equation}) over the CV and applying the Gauss's law with the incompressible condition:

\begin{equation}
\label{equ:mom_vol}
\partial_t \int_{CV} u_i dV + \oint_{A_{CV}} u_i \vec{u} \cdot d\vec{A} = - \int_{CV} \partial_i P dV + \int_{CV}TdV \delta_{iz} + \sqrt{\frac{Pr}{Ra}} \oint_{A_{CV}} (\nabla u_i) \cdot d\vec{A}
\end{equation}
\begin{equation}
\label{equ:therm_vol}
\partial_t \int_{CV} T dV + \oint_{A_{CV}} T \vec{u} \cdot d\vec{A} =  \sqrt{\frac{1}{RaPr}}\oint_{A_{CV}} (\nabla T) \cdot d\vec{A}
\end{equation}

Here $A_{CV}$ represents the enclosed area of the control volume. The convective and diffusive terms are now represented by the integral of flux. The physical variables considered in the simulation are the three components of velocity, temperature and pressure. The simplest way to locate the physical quantities is to use the collocated grid for all the physical variables. However, such grid arrangement may lead to odd-even decoupling between the velocity and the pressure fields. Therefore, our solution is to use the so-called staggered grid \citep{harlow1965numerical} instead of the collocated one. This strategy allows that the grid cells corresponding to the three velocity components $u, v, w$ to be shifted by half a grid cell in the $x, y, z$ directions respectively. Therefore, the control volumes for each components of velocity are shifted (staggered) as well. Such grid design is illustrated in fig. \ref{fig:stagger} (for simplicity, two-dimensional staggered grid is drawn). The interpolation scheme with fourth-order precision is constructed for evaluating face-averaged variables in our code.

The time integration of momentum and temperature equations are conducted by the so-called Euler-leapfrog scheme, which can be represented by 
\begin{equation}
\phi^{n+1} = \phi^{n-1} + 2\Delta t [C(\phi^n) + D(\phi^{n-1}) + B(\phi^n)]
\end{equation}
where $\phi$ is some physical quantity. The integer $n$ represents the time level and $\Delta t$ is the time step, and $C(...)$,  $B(...)$ and $D(...)$ corresponds to the convective terms, body force terms and diffusive terms. In such time-advance scheme, the convective terms and body force terms are temporally-discretized by the leapfrog scheme while the diffusive terms are discretized by the Euler forward scheme. However, such scheme exhibits phase error according to von Neumann stability analysis \citep{durran2013numerical}. To prevent the accumulation of phase errors, the averaging step of the results from the time levels $n-1$ and $n$ has been applied for every $N_p$ time steps:
\begin{equation}
\phi^{n+1} = 0.5\phi^{n-1} + 0.5\phi^{n} + 1.5\Delta t [C(\phi^n) + D(\phi^{n-1}) + B(\phi^n)]
\end{equation}
In our simulations, $N_p$ is chosen to be 50, which is determined by an unpublished stability test.

\subsection{Pressure solver}
The pressure term in the momentum equation is solved by the so-called Chorin projection method \citep{chorin1968moc}. The first step of this method is to solve the intermediate non-solenoidal velocity field $\vec{u}^{*,n+1}$ with the pressure gradient term being neglected
\begin{equation}
\label{momeqlessnop}
	\frac{\vec{u}^{*,n+1}-\vec{u}^n}{\Delta t} = -u_j \partial_j u_i + T \delta_{iz} + \sqrt{\frac{Pr}{Ra}}\partial_j\partial_j u_i
\end{equation}
Comparing Eq.\ref{momeqlessnop} with the original momentum equation,
\begin{equation}
\label{momeqlesswithp}
	\frac{\vec{u}^{n+1}-\vec{u}^n}{\Delta t} = -\partial_i P -u_j \partial_j u_i + T \delta_{iz} + \sqrt{\frac{Pr}{Ra}}\partial_j\partial_j u_i
\end{equation}
one can subtract Eq. \ref{momeqlesswithp} from Eq. \ref{momeqlessnop} and take divergence on both side to eliminate the terms related to $\vec{u}^n$ and $\vec{u}^{n+1}$ by applying to the incompressible condition. This gives the Poisson equation related to the pressure and the source terms originating from the non-solenoidal velocity field
\begin{equation}
\label{poiseq}
	\nabla^2 (P\Delta t) =  \nabla \cdot \vec{u}^{*,n+1}
\end{equation}
 Finally, the solenoidal velocity field $\vec{u}^{n+1}$ can be evaluated from the pressure gradient by
\begin{equation}
	\vec{u}^{n+1} = \vec{u}^{*,n+1} - \nabla (P\Delta t)
\end{equation}
The three dimensional Poisson equation for the pressure is solved by the separation of variables method \citep{kaczorowski2008nrnefm6}. This method makes use of eigen-decomposition to decouple the three-dimensional Poisson equations to several independent one-dimensional Poisson equations that can then be solved by simple tridiagonal matrix solver. The procedure of eigen-decomposition involves very massive matrix-to-vector multiplications that can be done efficiently by \textit{DGEMM} routine provided by Intel MKL. With the divergence-free velocity being updated, the heat equation will next be solved.


\begin{figure}[h!]
	\centering
	\includegraphics[width=.7\textwidth]{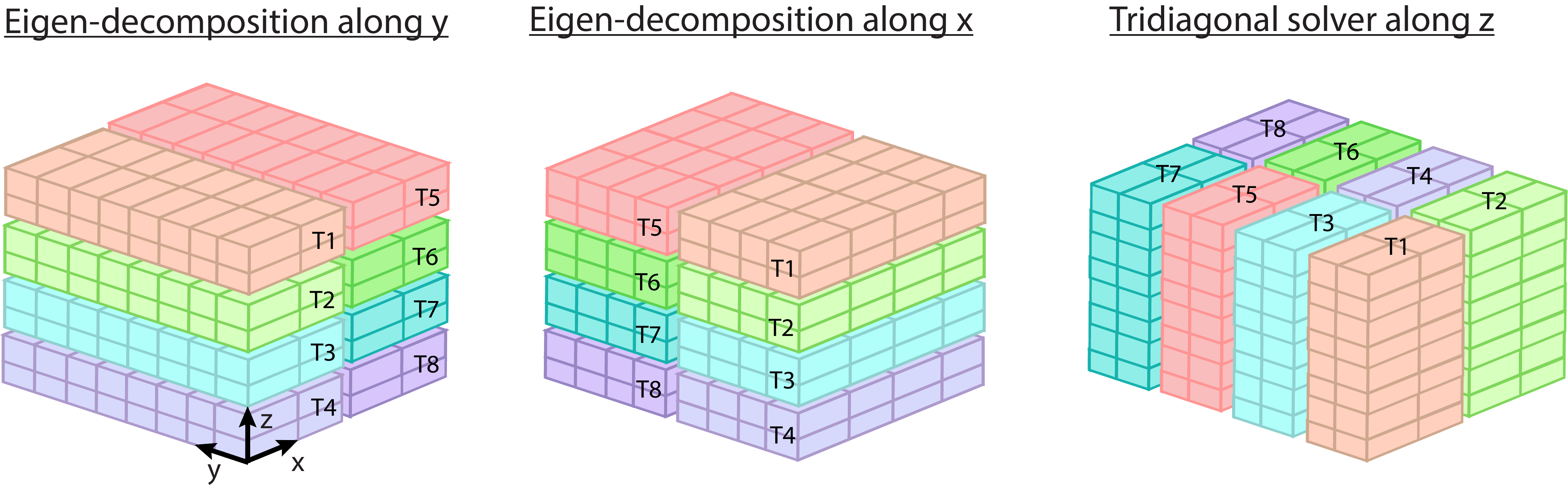}
	\caption{Illustration of the subdivision of domains for eigen-decomposition in the Poisson solver. The number T1 to T8 represents the index of MPI tasks.}
	\label{fig:pendiff}
\end{figure}

\subsection{Parallelization}
The \textit{CUPS} code makes use of pencil domain decomposition for parallel processing. Specifically, the domain is divided along the $x$ and $z$ directions while the data along the $y$ direction is kept intact. Each core only stores data from the pencil to which it belongs  during the computation, together with the additional two ghost grids on each side of each dimension which is for the fourth-order spatial interpolation. Therefore, after each time step, the inter-core swapping is needed to exchange data with neighbouring MPI task such that the data in ghost grids can be updated for calculation in the next time step.

\begin{figure}[h!]
	\centering
	\includegraphics[width=.7\textwidth]{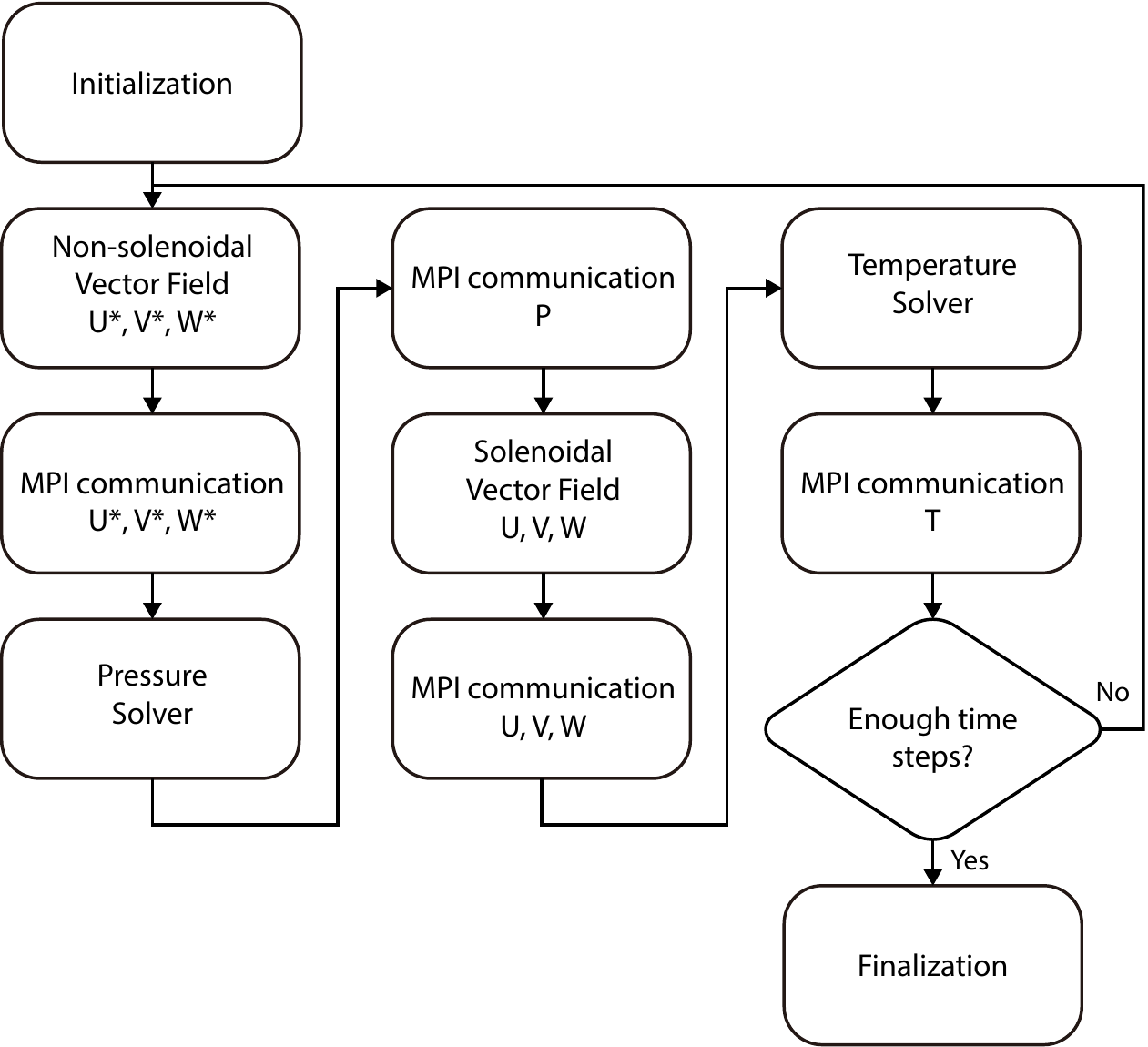}
	\caption{Flowchart of the single resolution scheme.}
	\label{fig:single res flowchart}
\end{figure}

The all-to-all communication is needed in the pressure solver. The eigen-decomposition is firstly done along the y direction where the y-oriented pencil suits for this purpose without additional data swapping. However, to continue the eigen-decomposition along x direction in the next step, data has to be exchanged such that the pencil becomes x-oriented. An example is shown in figure \ref{fig:pendiff} which demonstrates how the data exchange is done throughout the pressure solver. Next, the tridiagonal matrix solver along z direction requires z-oriented pencil which therefore needs another data swapping. Then the backward eigen-decompositions take further data swapping. 


To have better understanding of the flow of our algorithm written in single resolution version, we have constructed a flowchart as shown in Fig.\ref{fig:single res flowchart}. For more details of the \textit{CUPS} code, we refer to \citep{kaczorowski2013jfm}.
                                                                                                                                                                                                       

%% file: multires.tex
\section{Multiple resolution scheme}
The basic concept of multiple-resolution method is to solve the diffusion-advection equation and Navior-Stokes equations on different meshes that respectively resolves the Batchelor and Kolmogorov length scales. For diffusion-advection equation, it is solved in denser grid while for the momentum equation and the pressure solver, they are solved on coarser grid. Since the temperature and momentum are coupled, the velocity field has to be interpolated to match the denser mesh for temperature in every time step. However, there is constraint that the mass conservation of the newly interpolated velocity field is preserved. On the other hand, the temperature field has to be integrated to match the coarser mesh for velocity in every time step. We will show later that although the interpolation and integration procedures introduce additional computational costs, the overall performance of multiple-resolution code is superior to the single-resolution code owing to the fact that the momentum and the pressure solvers are the most time-consuming parts of the code.

The routine for the momentum and the temperature solvers in this multiple-resolution scheme is similar to the single resolution scheme. The only difference is the additional procedures for the downscale interpolation of velocity from coarse to dense mesh, and the upscale integration of temperature from dense to coarse mesh. In below, details of these procedures will be described. 

\subsection{Grid design}
In the multiple-resolution scheme, two sets of staggered grids are designed, denser one for temperature solver and coarser one for velocity (or pressure) solver. To simulate reliable results, the dense and coarse mesh have to resolve the Batchelor and Kolmogorov length scales respectively.
\begin{figure}[h!]
	\centering
	\includegraphics[width=.4\textwidth]{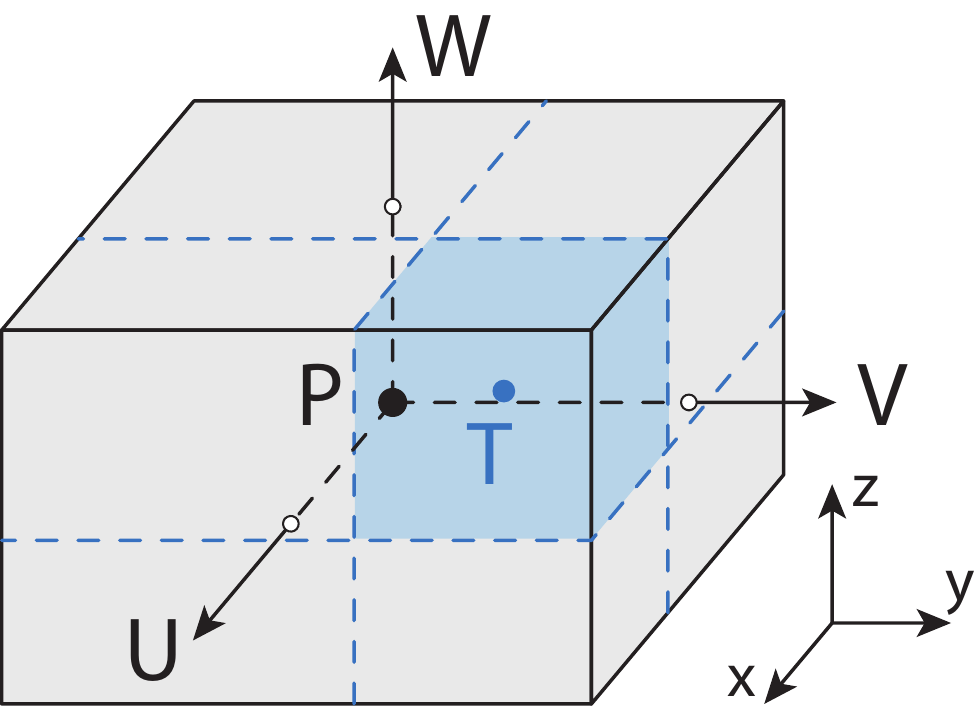}
	\caption{A sketch of the grid design in multiple-resolution scheme. Black solid grid denotes the coarse grid corresponding to the momentum solver (or pressure solver), and the blue dashed grid denotes the dense grid corresponding to the temperature solver. One of the dense grid cells is indicated by the volume filled blue.}
	\label{fig:grid_design}
\end{figure}
We use the refinement factors $M_x$, $M_y$ and $M_z$ to represent the number of dense grid points inside a single coarse grid cell along the $x$, $y$ and $z$ directions. To give an impression of the grid design, figure \ref{fig:grid_design} illustrates an example of how the dense and the coarse meshes are designed in the case of $M_x=M_y=M_z=2$. Under the framework of finite volume method, the velocity components $U$, $V$, and $W$ denoted in the figure represent the velocities averaged over the cell surface of coarse grid. The temperature grid point is placed at the center of the dense cell, which represents the volume-averaged temperature over the entire dense cell. As illustrated in fig.\ref{fig:grid_design}, the dense cells are not required to be equally divided from the coarse cell. For our multiple-resolution scheme, the procedure of generating the grid is as follows (take the $x$ direction as an example): Firstly, the dense grid is generated, then the positions of every $M_x$ dense grid is determined to be the coarse grid. The criteria on grid generation is similar to the single-resolution scheme, i.e. the grids has to be refined near the boundary layers to resolve the fine structures near the walls as proposed by Shishkina \textit{et al.} \citep{shishkina2010njp}.

\begin{figure}[h!]
	\centering
	\includegraphics[width=0.8\textwidth]{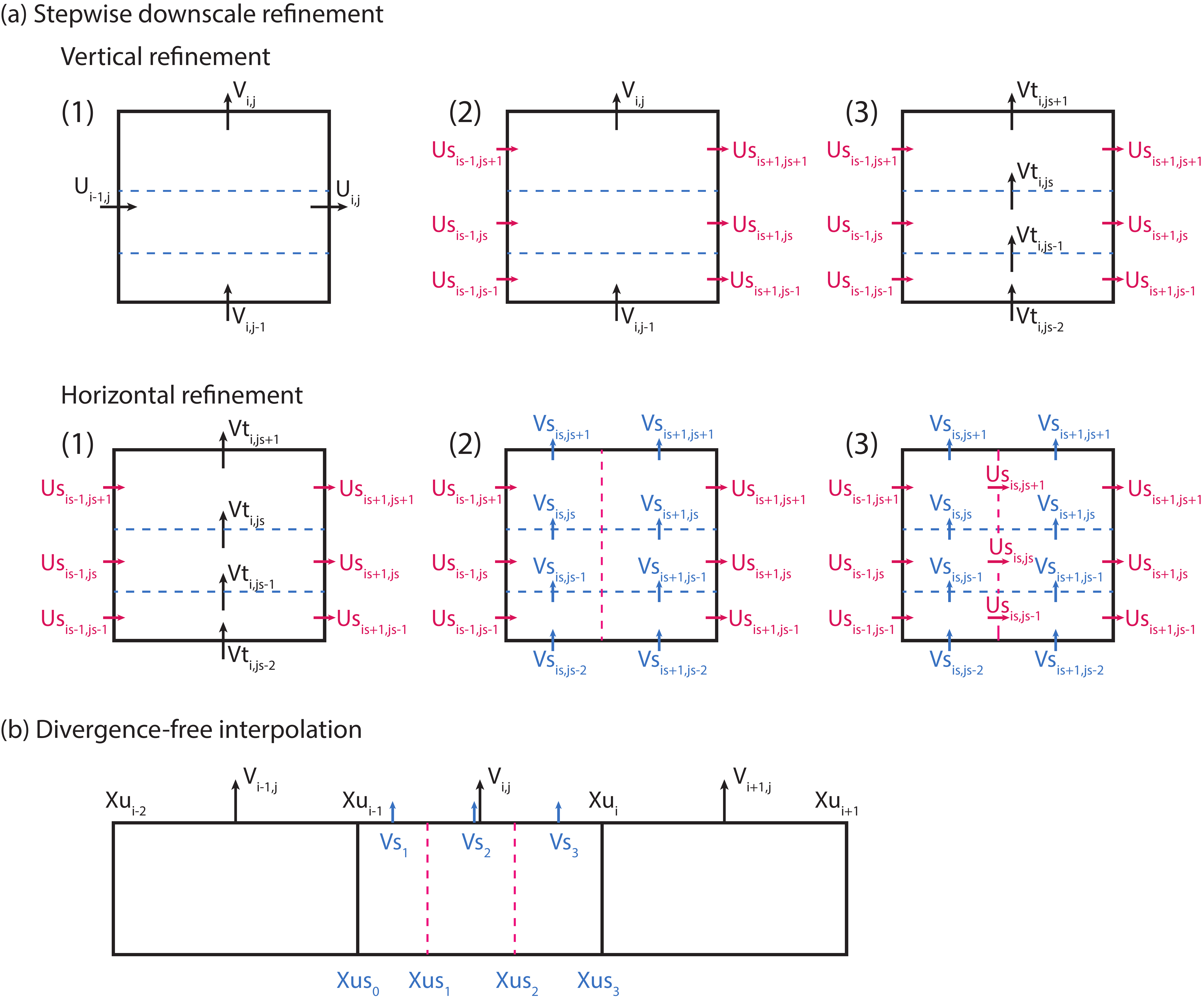}
	\caption{(a) Stepwise divergence-free refinement of the velocity field; and (b) a schematic of the divergence-free interpolation scheme.}
	\label{fig:stepwise}
\end{figure}

\subsection{Downscale refinement of the velocity field}
The multiple-resolution scheme involves interpolation of velocity field from coarse to dense grid for the use in the temperature solver. Here we propose a stepwise refinement scheme to reconstruct the divergence-free velocity field in the dense mesh which has previously been adopted in magnetoconvection simulation in which non-solenoidal magnetic field has to be ensured \cite{balsara2001jcp,li2004jcp}. For simplicity, we demonstrate the idea of stepwise refinement scheme in the two-dimensional case shown in fig. \ref{fig:stepwise} (a).

First of all, the outer wall normal velocity $U$ can be subdivided into several $Us$ through interpolation (the mass-conserving interpolation technique will be introduced soon). Next, the vertical velocity at the inner horizontal surface (illustrated by blue dashed line in fig. \ref{fig:stepwise} (a)) can be evaluated through the following procedure: Starting from the top sub-cell in fig.\ref{fig:stepwise}(a), the three velocity components $V_{i,j}$, $Us_{is-1,js+1}$ and $Us_{is+1,js+1}$ are known already, and thus the normal velocity at the inner surface $Vt_{i,js}$ can be obtained through the conservation constraint $Vt_{i,js}=-(Us_{is+1,js-1}+Us_{is+1,js+1}+V_{i,j})$. After obtaining the top inner surface velocity, the next inner surface velocity can be evaluated according to the same conservation constraint. Similar manner then applied to the rest of the inner surfaces.

After completing the vertical refinement, the evaluation will continue along the horizontal direction. As illustrated in fig.\ref{fig:stepwise} (a), the cell volume are subdivided into three cells vertically numbered as 1,2 and 3. Again, the vertical velocity $Vs$ at the outer surface of the cell is firstly evaluated by a mass-conserving interpolation. Once the outer surface velocities $Vs$ of cell 1s to 6s are obtained, the refined horizontal velocity at the vertical inner surface, which is denoted by red dashed line in fig.\ref{fig:stepwise} (a), can be determined by the conservation constraint.

As mentioned before, another key idea is to propose a mass-conserving interpolation that split the normal velocities into several parts. We demonstrate how such interpolation can be done for $V_{i,j}$ splitting into $Vs$ as shown in fig. \ref{fig:stepwise}. Here we apply a third order spatial interpolation to refine the velocity. The velocity distribution near the index $i,j$ can be written in the form of $f(x)=\sum_{3}^{n=1}\zeta_nX^{n-1}$. To determine the coefficients $\zeta_n$, we have applied the mass-conserving constraints given below:

\begin{equation}
\left\{\begin{matrix} 
V_{i-1,j}\Delta X_{i-1} &= \sum_{n=1}^{3}\zeta_n\int_{Xu_{i-2}}^{Xu_{i-1}}x^{n-1}dx
\\ V_{i,j}\Delta X_{i} &= \sum_{n=1}^{3}\zeta_n\int_{Xu_{i-1}}^{Xu_{i}}x^{n-1}dx
\\ V_{i+1,j}\Delta X_{i+1} &= \sum_{n=1}^{3}\zeta_n\int_{Xu_{i}}^{Xu_{i+1}}x^{n-1}dx
\end{matrix}\right.
\end{equation}
where $\Delta X_{i}=Xu_{i}-Xu_{i-1}$. The systems of equations can be further simplified as a linear equation $\vec{V}=\textbf{A}\vec{\zeta}$, where
\begin{equation}
\vec{V}=\left\lgroup V_{i-1,j}, V_{i,j}, V_{i+1,j}\right\rgroup^T,
\end{equation}
\begin{equation}
\vec{\zeta}=\left\lgroup \zeta_1,\zeta_2,\zeta_3\right\rgroup^T,
\end{equation}
and 
\begin{equation}
\textbf{A}=
\left\lgroup\begin{matrix} 
1 & \frac{1}{2}\left(Xu_{i-1}+Xu_{i-2}\right) &\frac{1}{3}\left(Xu^2_{i-1}+Xu_{i-1}Xu_{i-2}+Xu^2_{i-2}\right)\\
1 & \frac{1}{2}\left(Xu_{i}+Xu_{i-1}\right) &\frac{1}{3}\left(Xu^2_{i}+Xu_{i}Xu_{i-1}+Xu^2_{i-1}\right)\\
1 & \frac{1}{2}\left(Xu_{i+1}+Xu_{i}\right) &\frac{1}{3}\left(Xu^2_{i+1}+Xu_{i+1}Xu_{i}+Xu^2_{i}\right) 
\end{matrix}\right\rgroup.
\end{equation}
In such a way, the coefficients can be determined by $\vec{\zeta}=\textbf{A}^{-1}\vec{V}$. Thus, the velocity distribution $f(x)$ is known. Since the conservation law is applied as the constrains in the calculation of $\vec{\zeta}$, this velocity profile will automatically satisfy the divergence-free requirement. Then, the refined velocity $Vs$ can be determined directly by integrating this velocity distribution over their occupying grid size. We demonstrate such integration as follows (for the case of $Mx=3$): As the refined velocity $Vs$ stands for the average-velocity over the surface, it yields
\begin{equation}
\label{equ:refinement_int}
\left\{\begin{matrix} 
Vs_{1}\Delta Xs_{1} &= \sum_{n=1}^{3}\zeta_n\int_{Xus_{0}}^{Xus_{1}}x^{n-1}dx\\
Vs_{2}\Delta Xs_{2} &= \sum_{n=1}^{3}\zeta_n\int_{Xus_{1}}^{Xus_{2}}x^{n-1}dx\\
Vs_{3}\Delta Xs_{3} &= \sum_{n=1}^{3}\zeta_n\int_{Xus_{2}}^{Xus_{3}}x^{n-1}dx\\
...
\end{matrix}\right.
\end{equation}
where $\Delta Xs_i=Xus_i-Xus_{i-1}$. The refinement scheme Eq.\ref{equ:refinement_int} can also be written as a linear equation $\vec{V}_s=\textbf{A}_s\vec{\zeta}$, where
\begin{equation}
\vec{V}_s=\left\lgroup Vs_{1}, Vs_{2}, Vs_{3}, ...\right\rgroup^T,
\end{equation}
and 
\begin{equation}
\textbf{A}_s=
\left\lgroup\begin{matrix} 
1 & \frac{1}{2}\left(Xus_{1}+Xus_{0}\right) &\frac{1}{3}\left(Xus^2_{1}+Xus_{1}Xus_{0}+Xus^2_{0}\right)\\
1 & \frac{1}{2}\left(Xus_{2}+Xus_{1}\right) &\frac{1}{3}\left(Xus^2_{2}+Xus_{2}Xus_{1}+Xus^2_{1}\right)\\
1 & \frac{1}{2}\left(Xus_{3}+Xus_{2}\right) &\frac{1}{3}\left(Xus^2_{3}+Xus_{3}Xus_{2}+Xus^2_{2}\right)\\
... & ... & ...
\end{matrix}\right\rgroup.
\end{equation}
Notice that $\vec{\zeta}=\textbf{A}^{-1}\vec{V}$, then $Vs$ can determined by
\begin{equation}
\label{equ:refinement linear equation}
\vec{V}_s=\textbf{A}_s\left(\textbf{A}^{-1}\vec{V}\right)=\left(\textbf{A}_s\textbf{A}^{-1}\right)\vec{V}=\textbf{R}\vec{V}, 
\end{equation}
where $\textbf{R}=\textbf{A}_s\textbf{A}^{-1}$ is the $Mx\times3$ matrix. Note that the matrix $\textbf{R}$ is unchanged during the simulation as it depends on the grid spacing only, therefore the complicated calculation of $\textbf{R}$ will only be done in the beginning of the simulation. The mass-conserving interpolation given by Eq.\ref{equ:refinement linear equation} combined with the stepwise velocity refinement algorithm introduced just before completing the reconstruction of divergence-free velocity field from coarse to dense mesh. Also, we have checked that with this approach, the maximum divergence of the velocity field is as small as $10^{-17}$ which is the error of double precision order.

\begin{figure}[h!]
	\centering
	\includegraphics[width=0.9\textwidth]{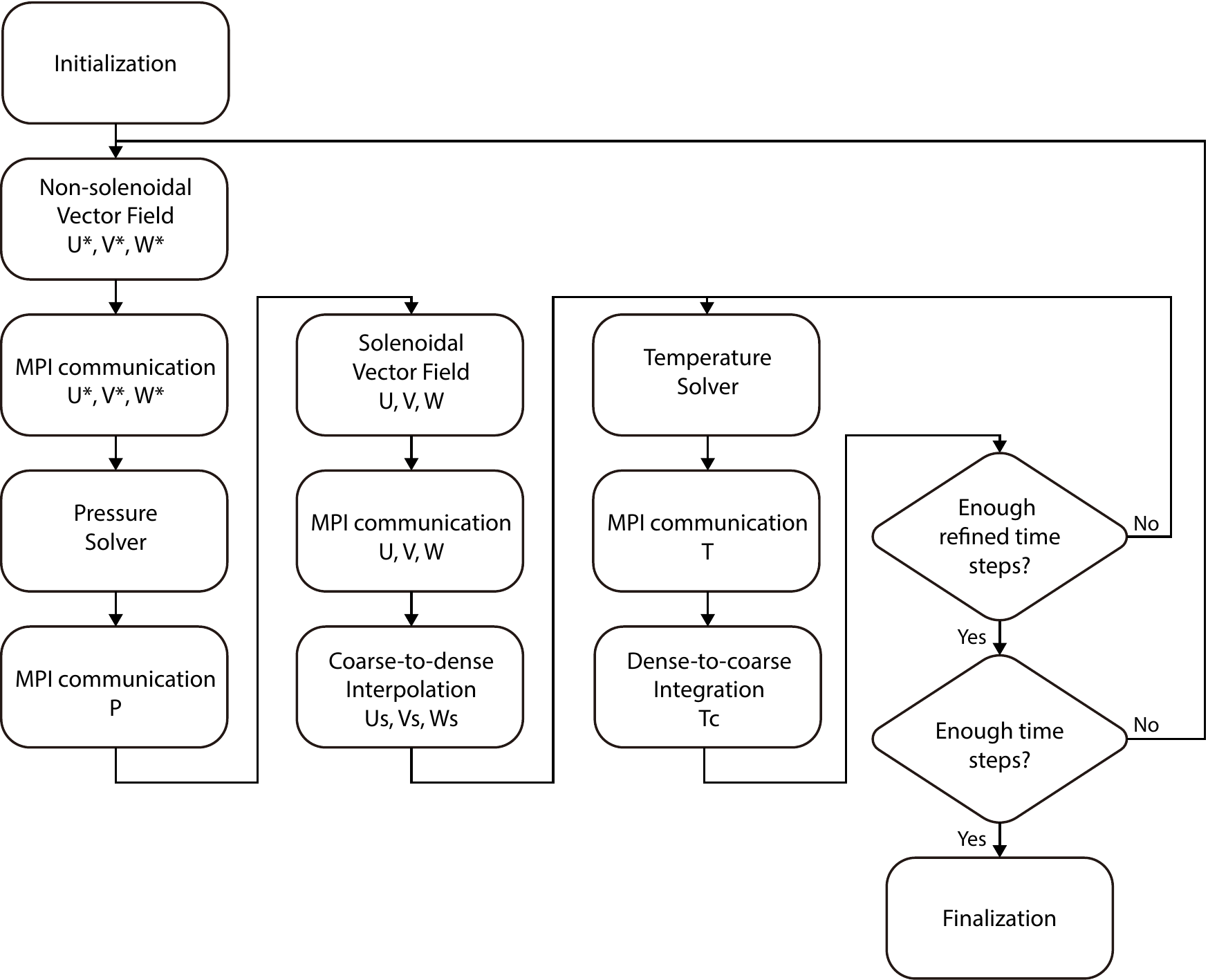}
	\caption{Flowchart of the multiple-resolution scheme.}
	\label{fig:multires_flowchart}
\end{figure}

\subsection{Upscale integration of temperature field}
For temperature field, an additional upscale integration of the temperature field is required to match the coarse velocity field. Note that in the finite-volume method, local temperature is defined as the average of temperature over the control volume, therefore it is trivial to calculate the temperature at the coarse grid by just summing up the values from different dense cells. A single coarse cell is composed of several dense cells, one can obtain the temperature at the coarse cell by a simple summation as follows
\begin{equation}
Tc(i,j,k) = \sum_{ks=1}^{M_z}\sum_{js=1}^{M_y}\sum_{is=1}^{M_x} T_{is,js,ks}\Delta xs_{is}\Delta ys_{is}\Delta zs_{ks}, 
\end{equation}
where $Tc$ represents the temperature of the coarse cell.

\subsection{Temporal strategy}
In the single-resolution code, the momentum and the temperature equations are solved with the same time step. However in multiple-resolution scheme, the computational cost could also be saved by using different time steps for different equations. The temporal strategy has been proposed by Ostillia-Monico \textit{et al.} \citep{ostillamonico2015jcp} in their finite-difference situation. With the same reason that coarser grid spacing allows for larger time steps according to Courant-Friedichs-Lewy criteria \citep{courant1928partiellen}, we have also implemented the interpolation of velocity field over time. Here we denote the temporal refinement factor as $M_t$, it means that for each step of momentum solver, the temperature solver is carried out by $M_t$ times. Under this temporal strategy, the intermediate velocity fields have to be computed and its procedure is as such: Momentum equation is firstly solved at every coarse time step, e.g. $U^{n-1}$ is updated to $U^{n}$ in a coarse time step. Then the intermediate velocity fields can be computed by linear interpolation:
\begin{equation}
U_i^{n-1}=U^{n-1}+\frac{i}{M_t}U^n.
\end{equation}
Since both $U^{n-1}$ and $U^{n}$ are divergence-free, the linear combination of $U^{n-1}$ and $U^{n}$ gives rise to the divergence-free velocity field as well. Thus, the mass-conserving condition can be retained throughout the entire simulation.

The routine of our multiple-resolution code can be summarized by a flowchart shown in fig.\ref{fig:multires_flowchart}. Comparing to the single-resolution code, our multiple-resolution version has additional inter-grid calculation which are the coarse-to-dense refinement of velocity and the dense-to-coarse integration of temperature.

%% file: validation.tex
\section{Verification and performance}
\subsection{Global heat flux}
Several simulations were conducted to verify the \textit{CUPS} code implemented with the multiple-resolution scheme. One of the verifications is to compare heat transport efficiency (characterized by the Nusselt number $Nu$) computed using different methods. There are several possible ways to compute $Nu$, one way is to directly evaluate the heat flux across each horizontal plane given by:

\begin{equation}
Nu(z)=(RaPr)^{1/2}\langle u_zT \rangle_{x,y,t} - \langle\partial_z T \rangle_{x,y,t}
\end{equation}
where $\langle ... \rangle_{x,y,t}$ denotes averaging over the horizontal plane and time. In principle, the heat flux across any plane is identical to each other provided that the averaging time is sufficiently long. Usually, the averaging of $Nu(z)$ along the $z$ direction is also taken to obtain a better estimation of convective heat flux which is given by $Nu=\sum Nu(z)/N_z$ where $N_z$ is the number of grid points along the $z$ direction. Besides direct evaluation, global Nu can also be computed from its exact relationships with the dissipation rates \citep{shraiman1990pra}:

\begin{equation}
Nu_T=\epsilon_T(RaPr)^{1/2}
\end{equation}
\begin{equation}
Nu_v=\epsilon_v(RaPr)^{1/2}+1.
\end{equation}
where $\epsilon_T$ and $\epsilon_v$ are the globally-averaged thermal dissipation rate and the viscous dissipation rate. In principle, these exact relationships hold true when the averaging time is sufficiently long. However, the dissipation rates are numerically approximated, in which the sharp velocity or temperature gradients may be missed if the grid points are inadequate. Therefore, by comparing the three values of $Nu$, one can obtain the first check of the multiple-resolution scheme on whether the dissipations have been well-resolved.

\begin{figure}[h!]
	\centering
	\includegraphics[width=1.0\textwidth]{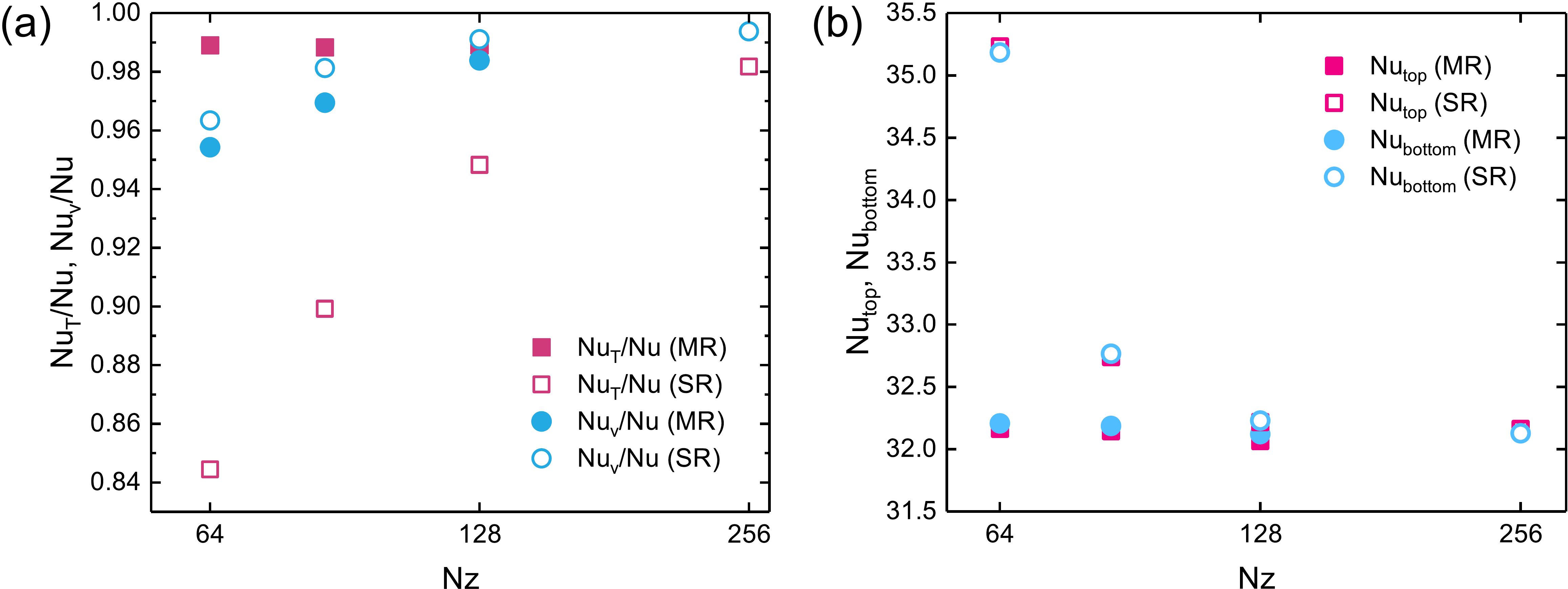}
	\caption{(a) Nusselt number computed from the exact relations for the energy and thermal dissipation rates, respctiveley. (b) Nusselt number computed from top and bottom plate temperature gradients  for $Ra=10^8$ and $Pr=4.38$. For multiple-resolution (MR), the dense mesh is fixed at $256^3$ and the coarse grid number $Nz$ is varied. For single-resolution (SR), the mesh is fixed at $Nz^3$.}
	\label{fig:Nufig}
\end{figure}

We have carried out different simulations for both multiple- and single-resolution codes with the parameters $Ra=10^8$ and $Pr=4.38$ in cubic geometry. All walls are set to be impermeable and no-slip. The sidewalls are adiabatic while the dimensionless temperature of the top and the bottom plates are set to be $-0.5$ and $0.5$, respectively. For the chosen parameters, one can estimate the Kolmogorov length scale to be about $l_k/H=8.9\times 10^{-3}$ and the Batchelor length scale to be about $l_b/H=4.2\times 10^{-3}$. In DNS, the grid size $\Delta$ is often chosen to be $\Delta/l_k \approx 1$ or $\Delta/l_b \approx 1$ \citep{shishkina2010njp}. Based on this criteria, the necessary grid points along a single dimension is about $112$ for velocity field and $238$ for temperature field. Here for single-resolution simulations, we began with the mesh $256^3$ which is thought as the adequate grid number for both temperature and velocity fields, and then lower the mesh successively to $64^3$. For multiple-resolution simulations, the dense mesh is fixed at $256^3$ such that the temperature field is always well-resolved while the coarse grid points is decreased from $128$ to $64$. Figure \ref{fig:Nufig} (a) shows $Nu_T/Nu$ and $Nu_v/Nu$ versus the number of grid points $Nz$ for both single-resolution (open symbols) and multiple-resolution simulations (closed symbols). First we focus on the results of single resolution. For $Nz=256$, which is believed to be the sufficient number, the differences between $Nu$ obtained from dissipations and $Nu$ directed evaluated from heat flux is below $2\%$, representing that both the velocity and temperature gradients are well-resolved. However, when $Nz$ decreases, both $Nu_\theta$ and $Nu_v$ deviates from $Nu$ but it is important to note that $Nu_\theta$ deviates much more significantly than $Nu_u$. It agrees with our expectation, as for large $Pr$ the temperature gradients are much sharper than the velocity gradients, and therefore the temperature field has stricter requirement on grid resolution. For multiple-resolution simulation, since the dense grid is fixed at $256^3$, the well-resolved temperature fields lead to the agreement between $Nu_\theta$ and $Nu$ for any coarse grid number. With coarse mesh reaching $128^3$, the velocity gradient can be resolved which therefore results in the agreement between $Nu_v$ and $Nu$. However, upon further decreasing in $Nz$, the difference between  $Nu_v$ and $Nu$ becomes apparent owing to the inadequate grid points to resolve the Kolmogorov length scale. Therefore, our multiple-resolution code can satisfy the exact relations whenever the Batchelor and Kolmogorov length scales are resolved by the dense and coarse grids separately.

Next, we compare Nusselt number computed from the top and bottom plate temperature gradients as shown in fig.\ref{fig:Nufig} (b). For single-resolution with the mesh $256^3$, the average of top and bottom $Nu$ is about $32.2$ and this number agrees excellently with the value from previous simulations conducted also in a cubic cell \citep{kaczorowski2013jfm,kaczorowski2014jfm}. Therefore, we regard it as the standard value. We observe that the top and bottom $Nu$ are very sensitive to the temperature mesh. As seen for the single-resolution scheme with the mesh fixed at $64^3$, the top and bottom $Nu$ are about 10\% larger than the standard value. However, for the multiple-resolution scheme with temperature mesh fixed at $256^3$ but varying the number of velocity grid point $Nz$ from $64$ to $128$, the top and bottom $Nu$ are within 2\% from the standard value.

\begin{figure}[h!]
	\centering
	\includegraphics[width=\textwidth]{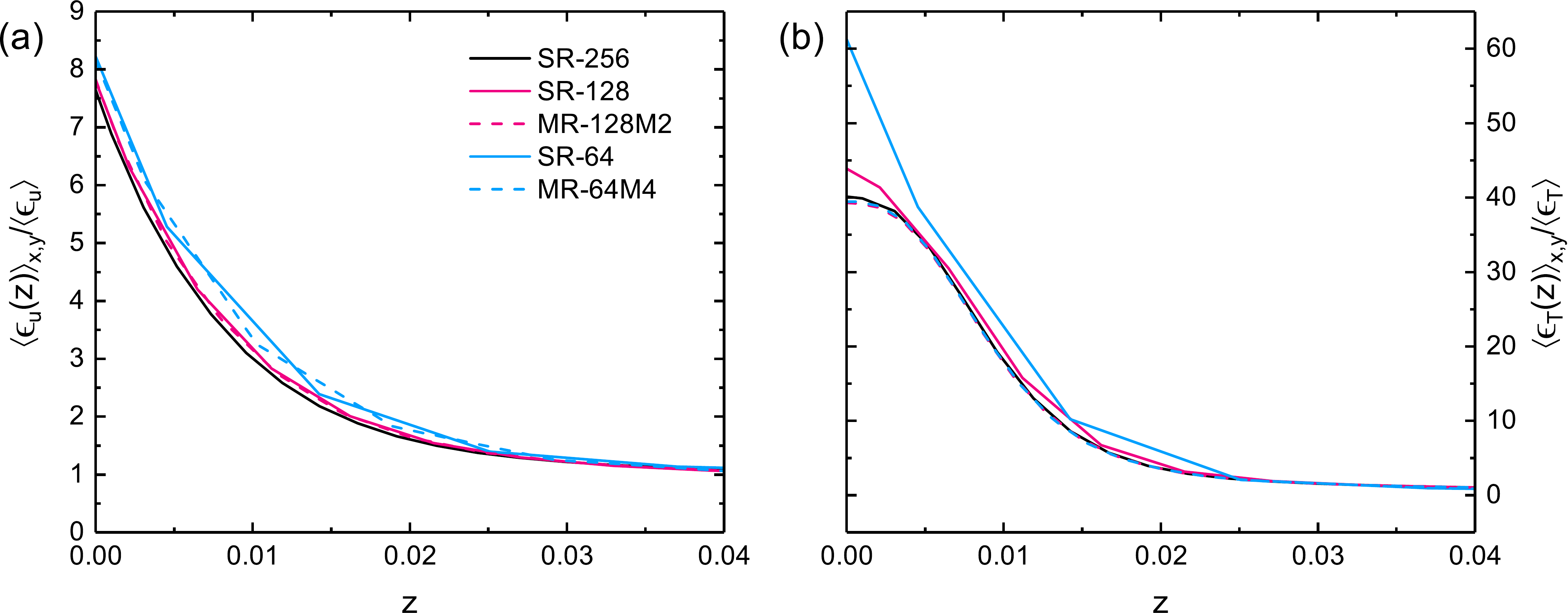}
	\caption{Horizontally averaged profiles of (a) viscous and (b) thermal dissipation rates at the bottom boundary layer normalized by the globally averaged viscous and thermal dissipation rates respectively. The legend code SR-256, for example, stands for the case simulated by single-resolution (SR) with mesh $256^3$, while MR-128M2 stands for the multiple-resolution (MR) with coarse mesh $128^3$ and refinement factor M=2.}
	\label{fig:dissipation_profile}
\end{figure}

For the parameters used in our test, the thermal and viscous dissipation rates are boundary layer-dominated. The deviations from the exact relations shown in fig.\ref{fig:Nufig} are mainly caused by the under-resolved dissipation rates near boundaries. In fig.\ref{fig:dissipation_profile} we examine the profile of the horizontally-averaged viscous and thermal dissipation rates near the bottom plane normalized by the globally-averaged ones as a function of height for both single- and multiple-resolution codes with different grid sizes. A well-resolved single-resolution simulation with grid size $256^3$ is chosen as a benchmark in this comparison. Results from the single-resolution code with $128^3$ (pink solid line, denoted as SR-128) and $64^3$ (blue solid line, denoted as SR-64) meshes, together with the multiple-resolution simulations using coarse mesh $128^3$ with refinement factor $M=2$ (pink dashed, denoted as MR-128M2), and coarse mesh $64^3$ with refinement factor $M=4$ (blue dashed, denoted as MR-64M4) are compared with the standard profile. In fig.\ref{fig:dissipation_profile} the normalized viscous dissipation is shown. For both SR-128 and MR-128M2 (pink curve), the normalized viscous dissipation profiles agree with the benchmark case. As for both SR-64 and MR-64M4 (blue curve), the profile is obviously under-resolved. This agrees with the results in fig.\ref{fig:Nufig}(a), that for $64^3$ mesh $Nu_u$ is about 4\% smaller than $Nu$. We next compare the normalized thermal dissipation profile shown in fig.\ref{fig:dissipation_profile}(b). For SR-128 and SR-64, the thermal dissipation rate near bottom plane is obviously under-resolved, clear deviations from the benchmark curve is observed. However, the multiple-resolution cases MR-128M2 and MR-64M4 can both successfully reproduce the well-resolved normalized thermal dissipation profile. This suggests that the thermal dissipation has stricter resolution requirement for $Pr>1$, especially near top and bottom planes where the thermal boundaries develop. Our multiple-resolution strategy therefore ensures that the temperature field is resolved, while in the same time avoids introducing redundant calculations in solving the velocity and the pressure.

\begin{figure}[h!]
	\centering
	\includegraphics[width=1.0\textwidth]{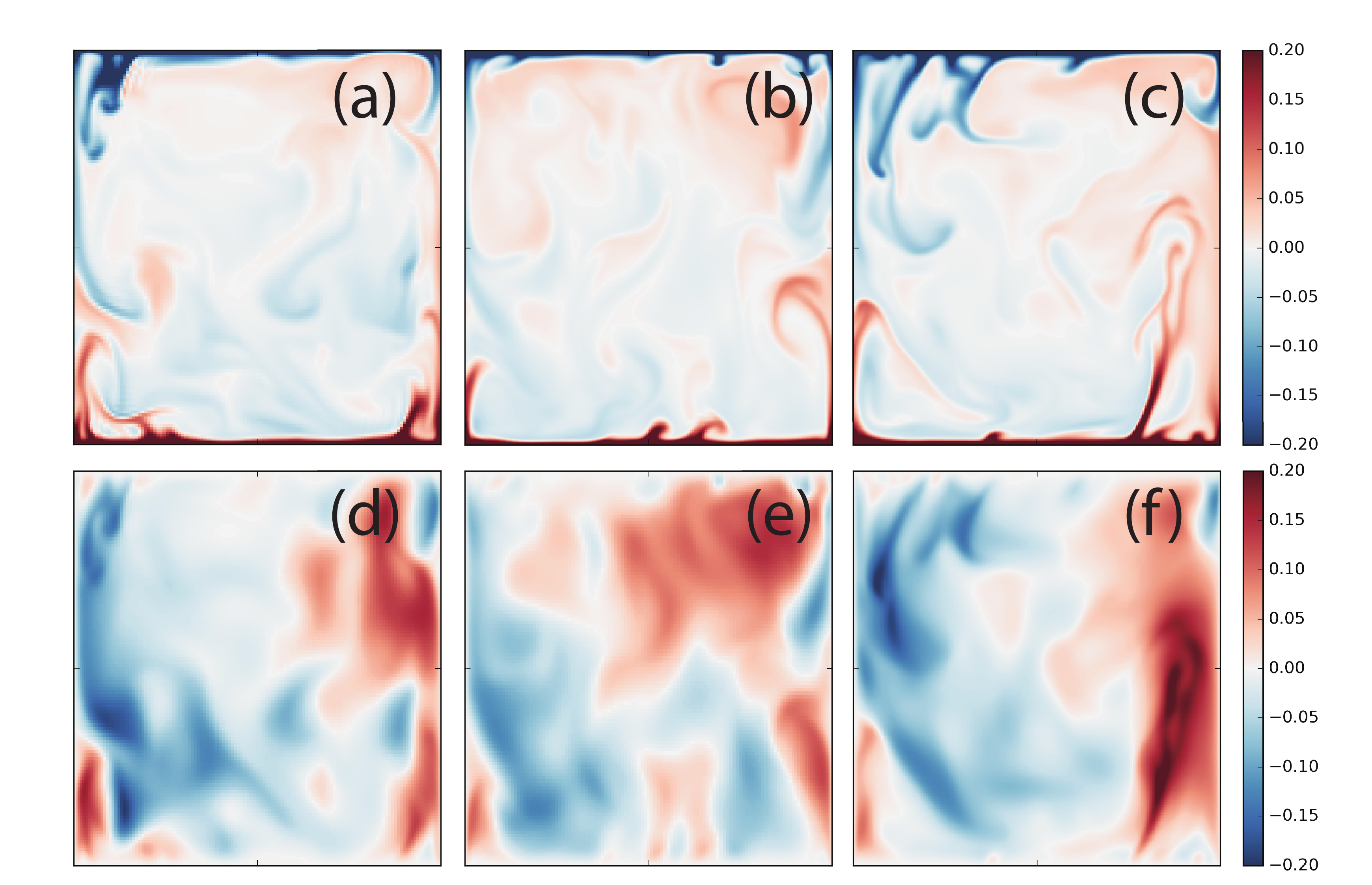}
	\caption{Vertical sections of instantaneous temperature (a, b, c) and vertical velocity fields (d, e, f). (a, d) are for the single resolution method with mesh $128^3$; (b, e) are for the multiple resolution method with coarse mesh $128^3$ and M=2; and (c,f) are for the single resolution method with mesh $256^3$.}
	\label{fig:T U field comparison}
\end{figure}

\subsection{Flow structures}
Besides comparing the global Nu, we also compare the flow structures obtained from different simulation runs with different grid resolutions by either single- or multiple-resolution schemes. Figure \ref{fig:T U field comparison} shows the temperature (top row) and vertical velocity (bottom row) snapshots obtained from different meshes. First of all, the instantaneous temperature fields display a typical thermal flow structures of Rayleigh-B\'enard convection where there are hot and cold thermal plumes in muchroom-like shape detaching from both the top and bottom boundary layers of the convection cell \citep{xi2004jfm}. These thermal plumes are actually steered by the large-scale circulation \citep{kadanoff2001pt}, which can be noticed from the vertical velocity snapshots. From the snapshots, one can observe that the temperature structures are much finer than the velocity structures owing to the fact that the viscous diffusion rate is larger than the thermal diffusion rate in our case of large $Pr$ ($> 1$). Now we compare how the grid resolution affects the ability to resolving thermal plumes. In fig.\ref{fig:T U field comparison} (a) where the mesh is $128^3$, the temperature field is under-resolved, as can be seen from the fuzzy edges of the thermal plumes. In contrast, with the larger mesh of $256^3$, the plume structures can be accurately resolved as shown in fig.\ref{fig:T U field comparison} (b,c). For vertical velocity structure, the mesh of $128^3$ as shown in fig.\ref{fig:T U field comparison} (d) and (e) can equally well resolve the velocity structures as the mesh of $256^3$ shown in fig.\ref{fig:T U field comparison} (f). Here with the multiple-resolution scheme, the tailored mesh can be adopted to suit for the thermal and velocity structures simultaneously and in the mean time saves the computational time, memory and snapshots storage.

\subsection{Small-scale properties}
\begin{figure}[h!]
	\centering
	\includegraphics[width=1.0\textwidth]{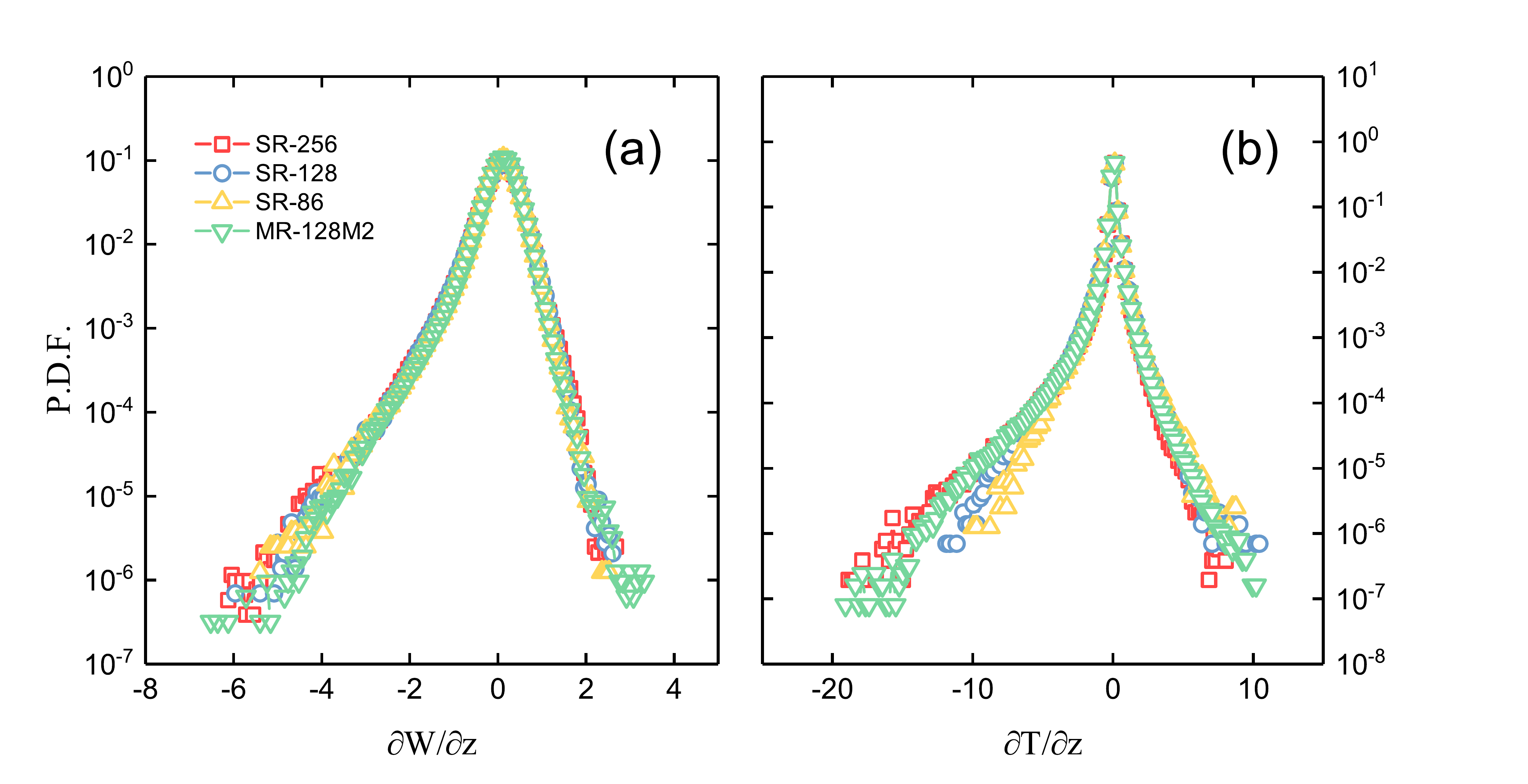}
	\caption{Probability density function (PDF) of (a) $\partial w/\partial z$ and (b) $\partial T/\partial z$. The legend code SR-256, for example, stands for the case simulated by single-resolution (SR) with mesh $256^3$, while MR-128M2 stands for the multiple-resolution (MR) with coarse mesh $128^3$ and refinement factor M=2.}
	\label{fig:pdf comparison}
\end{figure}

The local quantities can provide a stricter check than only the global heat flux and the overall flow structures. Fig.\ref{fig:pdf comparison} shows the probability density function (PDF) of the vertical velocity gradient $\partial_z W$ and the temperature gradient $\partial_z T$ at the center of the convection cell. In fig.\ref{fig:pdf comparison} (a), we observe that there is no significant differences between the velocity PDFs obtained from different resolutions from $86$ to $256$ grid points. It implies that the velocity gradients at the center can be well-resolved even at $86$ grid points. In contrast, the temperature PDFs are sensitive to the choice of resolution. First of all, for single-resolution with the mesh of $256^3$ or multiple resolution with coarse mesh of $128^3$ and refinement factor $2$, their PDFs are comparable to each other where the PDFs are skewed to the negative side due to the intermittent passage of thermal plumes. The agreement of the two PDFs suggests that multiple-resolution scheme is also suitable to capture the extreme temperature events once the resolution requirements are satisfied. However, all those extreme events cannot be well captured if one lowers the grid resolution for the temperature field. For example, using the single-resolution code with mesh size of $128^3$ or $86^3$, the tails of PDFs becomes less skewed, implying that some of the sharp temperature gradients are missed by those resolutions. 

More detailed descriptions of statistical quantities of the probability density function of $\partial_z W$ and $\partial_z T$ can be seen in Table \ref{tab:stat_quantities}. We examine the skewness defined $S=\frac{\langle (\phi-\langle \phi\rangle_t)^3\rangle_t}{\langle (\phi-\langle \phi\rangle_t)^2\rangle_t^{3/2}}$, which describes the asymmetricity of the data distribution, and kurtosis $K=\frac{\langle (\phi-\langle \phi\rangle_t)^4\rangle_t}{\langle (\phi-\langle \phi\rangle_t)^2\rangle_t^{2}}$, which measures the steepness of the data. A negative (positive) skewness implies that the data is right- (left-) skewed, and larger kurtosis stands for a steeper peak of the data distribution. For $\partial_z W$, there's not much difference between our test cases in skewness and kurtosis. However, for $\partial_z T$, significant difference between the resolved case SR-256 and the under-resolved cases SR-128 and SR-86 is observed. The absolute value of the skewness as well as the kurtosis of the resolved case is larger than the two under-resolved results. This can be understood by the fact that it would be less probable to capture the extreme events with insufficient resolution. As for the multiple-resolution method, both skewness and kurtosis agree with the revolved case. This results validates the accuracy of multiple-resolution scheme in small scale propeties of scalar turbulence simulation.

\begin{table}[]
	\centering
	\caption{Results for the small-scale quantities $\partial_z W$ and $\partial_z T$ using the single-resolution and multiple-resolution algorithms with different resolutions.}
	\label{tab:stat_quantities}
	\begin{tabular}{| l c c | c c | c c |}
				\hline
		\hline
		\multirow{2}{*}{ID}   &  \multirow{2}{*}{Temperature mesh}  & \multirow{2}{*}{Velocity mesh}  &\multicolumn{2}{|c|}{$\partial_z W$}      &\multicolumn{2}{|c|}{$\partial_z T$} \\
		 &  &  & Skewness & Kurtosis & Skewness & Kurtosis \\
		\hline
		\hline
		MR-128M2&$256^3$&$128^3$ 		&-1.2  &   9.7   & -5.2  &  100.6    \\  
		SR-256    &$256^3$&$256^3$        & -1.2 & 10.1  & -5.9  &  111.7   \\
		SR-128    &$128^3$&$128^3$         & -1.2 &  9.0   & -3.3  &  53.9    \\
	    SR-86      &$86^3$&$86^3$           &-1.3  &  10.1   & -1.9  &  41.9   \\

		\hline
		\hline
	\end{tabular}
\end{table}

\subsection{Code performance}
\begin{figure}[h!]
	\centering
	\includegraphics[width=1.0\textwidth]{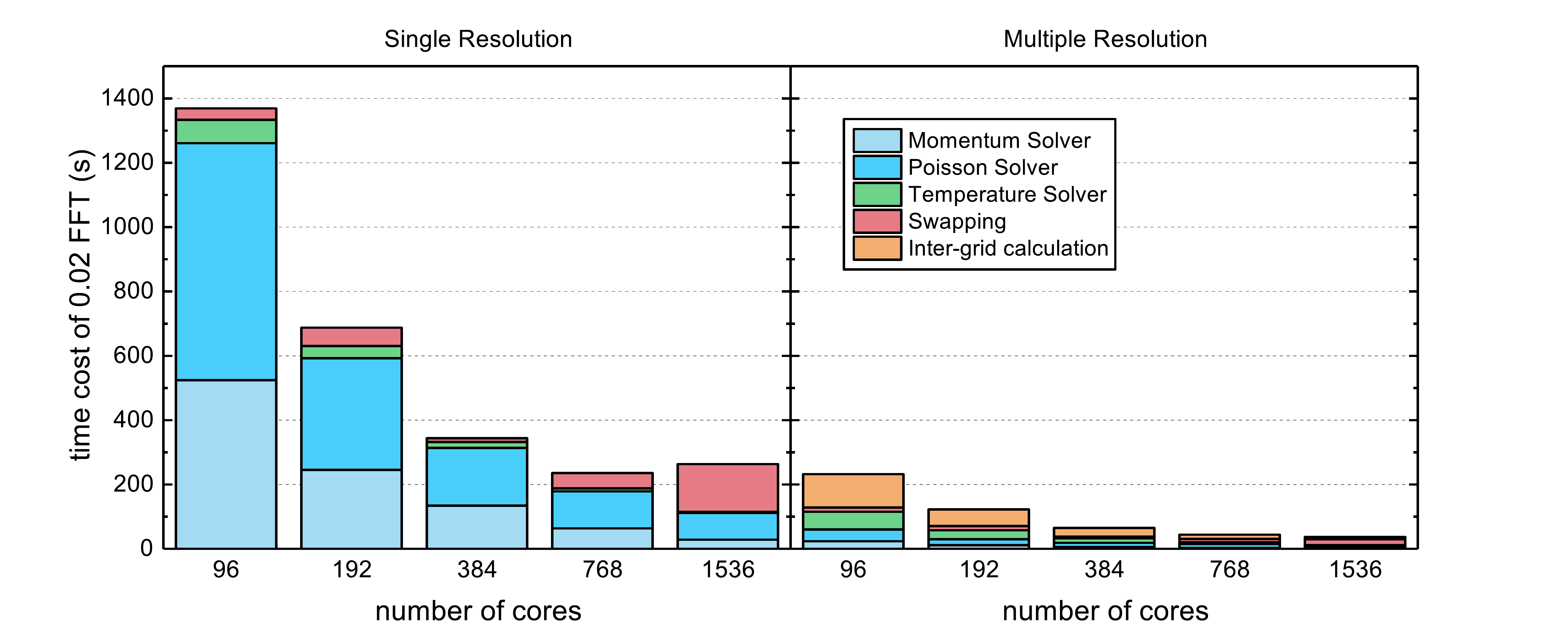}
	\caption{A time-cost test of the main routine for both the single- and the multiple-resolution schemes. For single-resolution, the mesh size used in this test is $1024^3$, and for multiple-resolution the coarse mesh size is $512^3$ and refinement factor for both space and time is $M=2$.}
	\label{fig:benchmark}
\end{figure}

We tested the code performance in Tianhe-2 from the National Supercomputer Center in Guangzhou. Each computing node installed with two 12-core Intel Xeon E5-2692 and each node contains 64GB memory. The performance profiling is done for both the single- and multiple-resolution codes. For single-resolution code, a $1024^3$ testing mesh has been used. For the multiple-resolution code, we chose a dense mesh with $1024^3$ grid points, while the refinement factor is $2$ for both space and time and it yields a $512^3$ coarse mesh. This refinement factor is suitable for $Pr=4.38$. Figure \ref{fig:benchmark} shows the distribution of the wall-clock time to finish 0.02 free-fall time unit for each solver routine as a function of core numbers. Both solver routines contain the momentum solver, Poisson solver, temperature solver and inter-core swapping. Inter-grid calculation is an additional routine for multiple-resolution scheme, which involves the coarse-to-dense interpolation of the velocity and dense-to-coarse integration of the temperature.  As the mesh size for the velocity solver and the Poisson solver is reduced to half, there is considerable reduction in computing time not only for the momentum solver and the Poisson solver, but also for the inter-core swapping. 

Over the tested range of core number, the multiple-resolution code is superior to the single-resolution code in spite of the additional inter-grid calculations. For example, with 96 cores, there is about five times speed up for multiple-resolution. The wall-clock time for 0.02 free-fall time is about 1350 seconds with the single-resolution code, whereas it is reduced to about 200 seconds for the multiple-resolution one. When 1536 cores are used, the speed-up for the single-resolution scheme is already vanished and it even takes more time than 768 cores, since the inter-core swapping becomes the bottleneck now. In contrast, for the multiple-resolution scheme, the code can still speed up for 1536 cores, though communication obviously starts to dominate in the total time cost. Obviously, the bottleneck caused by the communication time can be solved if one uses a supercomputer with more efficient interconnections. 

%% file: conclusion.tex
\section{Conclusion}
In summary, we have developed a multiple-resolution algorithm implemented with a finite-volume simulation code for turbulent flows with active or passive scalar fields. In the case of large Schmidt number (or Prandtl number), the momentum equations are solved in a coarse grid while the diffusion-advection equation for the scalar field is solved in a dense grid. Since velocity and scalar fields are computed with different meshes, comparing to the single-resolution scheme, additional steps are needed for the coarse-to-dense refinement of the velocity field and the dense-to-coarse integration of the temperature. We also developed the stepwise downscale refinement combined with third order spatial interpolation to reconstruct the divergence-free velocity in dense grid. In addition, the refinement can also be done temporally where the intermediate velocity field can be obtained through the linear interpolation in time.

We have verified the multiple-resolution code using the Rayleigh-B\'enard convection as an example. It is found that the multiple-resolution code can accurately reproduce the global heat flux, characterized by the Nusselt number, when both the velocity and temperature gradients are adequately captured by their own meshes.  Furthermore, the multiple-resolution code can also produce correctly the dissipation profiles, the velocity and temperature flow structures. Furthermore, besides the global heat flux and the general flow structure, the code can also capture correctly the local probability density function of velocity and temperature gradients at the center of the cell. All the results produced by the multiple-resolution scheme agrees with the traditional single-resolution simulation with full resolution. However, the computational resources required for the multiple-resolution scheme is much less than the single resolution one.

We have further tested the speed-up performance and the profiling of each routine of the code using the mesh of $1024^3$. These tests were carried out in Tianhe-2 in the National Supercomputer Center in Guangzhou. It was found that there is a significant speed-up for the multiple-resolution code; for example, the code is about five times faster than the single-resolution one when using 96 cores. However, the communication costs imposes a bottleneck to the speed-up when a larger number of cores is used.

\section{Acknowledgements}
This work was supported by the Hong Kong Research Grants Council (grant no. CUHK14301115 and CUHK14302317) and a NSFC$/$RGC Joint Research Grant N\_CUHK437$/$15; and by a Hong Kong PhD Fellowship.